\documentclass[epj]{svjour}

\usepackage{graphics}
\usepackage{amsfonts}
\newcommand{\ps}{p^\text{s}}

\newcommand{\eps}{\varepsilon}

\def\xa{x(\tau)}

\def\la{\lambda(\tau)}
\def\l{\lambda}

\def\beq{\begin{equation}}
\def\ee{\end{equation}}
\def\bi{\begin {itemize}}
\def\ei{\end{itemize}}

\def\eps{\epsilon}

\def\xa{x(\tau)}

\def\xxx{[x(\tau)|x_0]}

\def\la{\lambda(\tau)}

\def\l{\lambda}

\def\lsa{\tilde\lambda(\tau)}
\def\lt{\lambda_t}

\def\dst{\dot s_{\rm tot}(\tau)}

\def\dsm{\dot s_{\rm m}(\tau)}
\def\m{_{\rm m}}
\def\t{_{\rm tot}}

\def\hk {_{\rm hk}}
\def\ex{_{\rm ex}}

\def\pnt{p_{n(\tau)}(\tau)}

\def\pjp{p_{n_j^+}}
\def\pjm{p_{n_j^-}}
\def\wpm{w_{n_j^+n_j^-}}
\def\wmp{w_{n_j^-n_j^+}}
\def\njp{{n_j}^+}
\def\njm{{n_j}^-}

\def\tot{_{\rm tot}}
\def\m{_{\rm m}}

\def\F{{\cal F}}

\def\lo{\lambda_0}

\def\pst{p(\Delta s\t)}
\def\pmst{p(-\Delta s\t)}

\def\wnm{w_{nm}}
\def\wmn{w_{mn}}

\def\Sd{Stochastic thermodynamics~}
\def\sd{stochastic thermodynamics~}

\begin{document}
\title{Stochastic thermodynamics: Principles and perspectives}
\author{Udo Seifert 
}                     
%
%
\institute{   {II.} Institut f\"ur Theoretische Physik, Universit\"at Stuttgart,
  70550 Stuttgart, Germany }
\date{Received: date / Revised version: date}
%
\abstract{
Stochastic thermodynamics provides a framework for describing small systems
like colloids or biomolecules driven out of equilibrium but still in contact 
with a heat bath.
Both, a first-law like energy balance involving exchanged heat
and entropy production entering refinements of the second law
can consistently be defined along single stochastic
trajectories. Various exact relations involving the distribution of such
quantities like integral and detailed fluctuation theorems for total entropy
production and the Jarzynski relation follow from such an approach
based on Langevin dynamics. Analogues of these relations can be proven for
any system obeying a stochastic master equation like, in particular,
(bio)chemically driven enzyms or whole reaction networks. The
perspective of investigating  such relations  for stochastic field equations
like the Kardar-Parisi-Zhang equation is sketched as well.  
} 
\maketitle
\section{Introduction}
\label{intro}

\Sd as a conceptual framework combines the stochastic energetics approach
introduced a decade ago by Sekimoto \cite{seki98} 
with the idea that entropy can
consistently be assigned to a single fluctuating trajectory \cite{seif05a}. 
Stochastic
energetics has been developed for mesoscopic systems like colloidal
particles or single (bio)molecules driven out of equilibrium by 
time-dependent forces but still embedded in a heat bath of well-defined 
temperature. The dynamics of the few degrees of freedom are modelled by a
Langevin equation where the surrounding heat bath provides the thermal
noise. A first-law-like energy balance relates applied or extracted work, 
exchanged heat and 
changes in internal energy along the single fluctuating trajectory. In the
ensemble, these quantities acquire no longer sharp values but distributions.
Since dissipated heat is typically associated with changes in entropy
this program of using the classical thermodynamic notions to describe the
trajectories of these small systems obviously requires also the concept of 
a fluctuating entropy with corresponding distribution in the ensemble.

\Sd being based on stochastic equations of motion allows also to embed and
categorize two recent developments orginally derived independently using 
different equations of motion. First, the (detailed) fluctuation theorem
dealing with non-equilibrium steady states provides a symmetry between
the probability to observe asymptotically a certain entropy
production and the probability for the corresponding entropy annihilation
\cite{evan93,gall95,kurc98,lebo99}.
Second, the Jarzynski relation expresses the free energy difference
between two equilibrium states as a non-linear average over the
non-equilibirum work required to drive the system from one state to the other 
in a finite time \cite{jarz97,jarz97a,croo99,croo00}. 
Within \sd both of these relations can easily be derived
and the latter shown to be a special case of a
more general relation \cite{seif05a}.

The purpose of this article is to describe the principles of \sd in
a systematic way and to sketch a few perspectives. By character mainly a 
concise  review (with a 
few original results) 
from a personal unifying perspective,  no 
attempt is made to achieve a comprehensive historical presentation. Several 
(mostly review) 
articles can  provide complementary and occasionally broader perspectives
\cite{evan02,astu02,voll02,parr02,maes03,andr04,bust05,regu05,rito06,qian06,impa07a,harr07,zia07,kawa07,rond07}.
Focussing on the theoretical concepts I will point to experimental work
only where these experiments serve as an illustration of the theory.
Refererence to the growing number of  experiments and numerical 
simulations where these concepts are used to learn details about 
particular systems can also be found in those articles.

\section{Stochastic energetics}

\subsection{Stochastic Langevin dynamics} 

As  basic model, we consider overdamped motion $x(\tau)$
 of a ``particle''
or ``system''  governed by the Langevin
equation 
\beq
\dot x = \mu F(x,\l) + \zeta
\ee
where $F(x,\l)$ is a systematic force and $\zeta$ thermal noise
with correlation
$
\langle \zeta(\tau)\zeta(\tau')\rangle = 2 D \delta(\tau-\tau')
$
where $D$ is the diffusion constant. In equilibrium, $D$ and 
the mobility $\mu$ 
are related
by the Einstein relation
$
D=T \mu
$ where $T$ is the temperature of the surrounding medium
with Boltzmann's constant set to unity throughout the
paper to make entropy dimensionless. 
In stochastic
thermodynamics, on assumes that
the strength of the noise is not affected by the presence of
 a time-dependent force. The range of validity of this 
crucial assumption  can  be
tested experimentally or in simulations by comparing with 
theoretical result derived on the basis of this assumption.

The force 
\beq
F(x,\l)=-\partial_xV(x,\l) + f(x,\l) 
\ee
 can arise from a conservative potential $V(x,\l)$
 and/or be
applied to the particle directly as $ f(x,\l)$. Both sources
may be time-dependent through an external control
parameter $\la$ varied according to some prescribed 
experimental protocol from $\l(0)\equiv \l_0$ to 
$\l(t)\equiv \l_t$.

To keep the notation simple, we treat the coordinate $x$ as
if it were a single degree of freedom. In fact, all results
discussed in the following hold for an arbitrary number of
coupled degrees of freedom for which $x$ and $F$ become vectors
and $D$ and $\mu$ (possibly $x$-dependent)
matrices.

Equivalent to the Langevin equation is the 
 corresponding Fokker-Planck equation  
for the probability $p(x,\tau)$ to find the particle at $x$ at time 
$\tau$
as 
\begin{eqnarray}
\partial_\tau p(x,\tau)&=& - \partial_x j(x,\tau) 
\nonumber
\\
&=&-\partial_x \left(\mu F(x,\l)p(x,\tau) -D\partial_xp(x,\tau)\right) 
\label{eq:fp}
\end{eqnarray}
where $j(x,\tau)$ is the current.
This partial differential equation must be augmented by a
normalized initial
distribution $p(x,0)\equiv p_0(x)$. It will become crucial to distinguish
the dynamical solution $p(x,\tau)$ of this Fokker-Planck equation,
which depends on this given initial condition, from the solution $p^s(x,\l)$ 
for which the rhs of eq. (\ref{eq:fp})
vanishes at any fixed $\l$. The latter corresponds either to 
a steady state for a non-vanishing non-conservative force $f\neq0 $ 
or to  equilibrium for $f=0$, respectively. 

A third equivalent description of the dynamics is given by 
assigning a weight
\beq
p\xxx=\exp\left[-\int_0^td\tau[(\dot x -\mu F)^2/4D + \mu F'/2]\right]
\label{eq:PI}
\ee
to each path or trajectory. The last term in the exponent
arises from the Jacobian
$|\partial \zeta/\partial x|$ in the Stratonovitch discretization
required because the weight for a noise history $p[\zeta(\tau)]$
 is transformed into
the weight of a trajectory $p [x(\tau)]$.
Path dependent observables can then
be averaged using this weight in a path integral which requires
a path-independent normalization such that summing the weight
(\ref{eq:PI}) over all paths
is 1.

\subsection{First law}

As a major ingredient to stochastic thermodynamics, the first-law-like
energy balance 
\beq
dw=dV+dq
\ee
identified for a  Langevin equation by Sekimoto within his
stochastic energetics approach \cite{seki98} needs to be recalled.

The increment in work applied to the system
\beq
dw= (\partial V/ \partial \lambda) ~\dot \lambda ~d\tau + f ~ dx 
\label{eq:work}
\ee
consists of two contributions. The first term arises from
changing the potential (at fixed particle position) and the second
from applying a non-conservative force to the particle directly.
If one accepts these quite natural definitions, 
for the first law to hold along a trajectory the heat dissipated into the
medium must be 
identified with
\beq
dq= F dx  .
\ee
This relation is quite physical since in an overdamped system
the total force times the displacement corresponds to  dissipation. 
Integrated along a trajectory of given length, one obtains the
expression

\beq
 w[x(\tau)]= q[x(\tau)] + \Delta V= q[x(\tau)] + V(x_t,\lt)-V(x_0,\lo)
\label{eq:fl}
\ee

It is interesting and crucial to note that the heat dissipated along a trajectory
can also be written using
the path integral weight as

\beq
q[x(\tau)]= 
\int_0^tF(x,\tau)\dot x ~d\tau = T 
 \ln {p\xxx\over\tilde  p[\tilde x(\tau)|\tilde x_0]}  
\label{eq:heat} ,
\ee
where $\tilde p[\tilde x(\tau)|\tilde x_0]$ is the weight of the
backward path 
$
\tilde x(\tau)\equiv x(t-\tau)  
$
starting at $\tilde x(0) = x(t)$
under the time-reversed protocol
$
\lsa\equiv \l(t-\tau) .
$

In a recent experiment \cite{blic06}, 
the three quantities applied work, exchanged heat and 
internal energy were inferred from the trajectory of a colloidal particle
pushed periodically by a laser trap against a repulsive substrate. 
The measured non-Gaussian distribution for the applied work shows that
this sytem is driven beyond the linear response regime since it has been
proven that within the linear response regime the work distribution is always 
Gaussian \cite{spec04}. 
Moreover, the good agreement between the experimentally measured 
distribution and the theoretically calculated one indicates that the
assumption of noise correlations being unaffected by the driving
 is still valid in this regime
beyond linear response.

\section{Entropy production}

\subsection {Entropy production in the medium}

The heat dissipated into the environment should be identified with an
increase in entropy of the medium
\beq
\Delta s\m[x(\tau)] \equiv q[x(\tau)]/T.
\label{eq:dsm}
\ee From a purely classical perspective, it might be
disturbing to define an entropy change which depends on the trajectory
and thus becomes a stochastic quantity. Usually, entropy is considered
to be an ensemble quantity. In fact, it turns out that one can
and should go even one step further and assign a second contribution to entropy
as follows.

\subsection {System entropy}

Having solved the Fokker-Planck equation (\ref{eq:fp}), one defines as a trajectory
dependent entropy of the system the quantity \cite{seif05a}
\beq
s(\tau) \equiv  -\ln p(x(\tau),\tau)
\label{eq:s}
\ee
where the probability $p(x,\tau)$ obtained by solving the
Fokker-Planck equation is evaluated along the stochastic 
trajectory $x(\tau)$. Obviously, for any given trajectory
$\xa$, the entropy $s(\tau)$  depends on the given initial
data $p_0(x)$ and thus contains information on the
whole ensemble.
 What at first sight might look just as a definition 
becomes useful and interesting
through the following observations:

(i) Upon averaging with the given ensemble $p(x,\tau)$, the
trajectory-dependent entropy becomes the usual ensemble
entropy
\beq
S(\tau) \equiv  - \int dx ~p(x,\tau) \ln p(x,\tau) 
= \langle s(\tau)\rangle .
\label{eq:ens}
\ee

(ii)
In equilibrium, i.e. for $f\equiv 0$ and constant $\l$, 
this definition assigns a stochastic
entropy 
\beq
s(x(\tau))=(V(x(\tau),\l)-\F(\l))/T  ,
\ee with the free energy 
\beq
\F(\l)\equiv -T\ln \int 
dx
~\exp[-V(x,\l)/T].
\ee Thus, in equilibrium the well-known 
thermodynamic
relation now holds along the trajectory at any time.

(iii) Most significantly, the total entropy change
along a trajectory
\beq
\Delta s\t \equiv \Delta s\m + \Delta s
\ee 
with
\beq \Delta s= -\ln p(x_t,\lt)+\ln p(x_0,\lo)
\ee
now obeys an integral fluctuation theorem (IFT) \cite{seif05a}
\beq
\langle e^{-\Delta s\tot} \rangle =1  
\label{eq:R3}
\ee
which can be interpreted as a refinement of the second law
$\langle \Delta s\t \rangle \geq 0$. The latter follows from
(\ref{eq:R3}) by Jensen's inequality. Here and throughout
the paper the brackets $\langle ... \rangle$
denote the non-equilibrium average generated by the Langevin
dynamics from some given initial distribution $p(x,0)=p_0(x)$.

This integral fluctuation theorem for $\Delta s\t$ is quite universal
since it holds  for any kind of initial condition
(not only for $p_0(x_0)=p^s(x_0,\lambda_0)$),
any  time-dependence of force and potential, with (for $f=0$) and
without (for $f\not = 0$) detailed balance at fixed $\l$, and  any  length
of trajectory $t$.

\subsection {Invariance under coordinate transformations}

The entropy as defined in (\ref{eq:s}) has the formal deficiency that
strictly speaking $\ln p(\xa,\tau)$ is not defined since $p(x,\tau)$ is a
density. Apparently more disturbingly, this expression is not invariant under non-linear 
transformations of the coordinates. In fact, both deficiencies which also
hold for the ensemble entropy (\ref{eq:ens}) are  related
and can be cured easily as follows by implicitly
invoking the notion of relative entropy.

\def\ri{\{x_i\}}
\def\rit{\{x_i(\tau)\}}
\def\pi{\{p_i\}}
\def\pit{\{p_i(\tau)\}}

\def\ldb{\lambda_T}
\def\yi{\{y_i\}}
\def\yit{\{y_i(\tau)\}}

A formally proper definition of the stochastic entropy 
starts by describing the trajectory using canonical
variables. After integrating out the momenta, 
for a system with $N$ particles with Cartesian positions $\ri$,
one should define the entropy as
\beq
s(\rit) \equiv - \ln[ p(\rit)\ldb^{3N}]
\ee
where $\ldb$ is the thermal de Broglie length. If one now considers this
dynamics in other coordinates $\yi$, one should use
\beq
s(\yit) \equiv   - \ln[ p(\yit,\tau)\det \{\partial y/\partial x\}\ldb^{3N}
].
\ee
This correction with the Jacobian ensures that the entropy both on the
trajectory as well as on the ensemble level is independent of the coordinates
used to describe the stochastic motion. Of course, this statement is no
longer true  if the transformation from $\ri$ to $\yi$ is
not one to one. Indeed, if some degrees of freedom are integrated out
the entropy does and should change.

\section{A unifying integral fluctuation theorem (IFT)}

The IFT for entropy production  (\ref{eq:R3})
follows from a more general fluctuation theorem which unifies several
 relations previously derived independently. Based on the concept of
 time-reversed trajectories and time-reversed protocol \cite{kurc98,croo00,maes03},
it is easy to  prove the relation \cite{seif05a}
\beq
\langle \exp[-\Delta s\m] ~ p_1(x_t)/p_0(x_0)\rangle = 1
\label{eq:IFT}
\ee
for any  function $p_1(x)$ with normalization $\int dx~ p_1(x) =1$. 
Here, the initial distribution
$p_0(x)$ is arbitrary. By using the first law (\ref{eq:fl}) this
relation can also be written in the form
\beq
\langle \exp[-(w-\Delta V)/T]~ p_1(x_t)/p_0(x_0)\rangle = 1
\label{eq:IFT2}
\ee with no reference to an entropy change.
 
The arguably most natural choice for the function $p_1(x)$ is to
take the solution $p(x,\tau)$ of the Fokker-Planck equation at time $t$ which
leads to the IFT     (\ref{eq:R3})                     for the total entropy
production. Other choices lead to the following relations 
originally derived differently.

\subsection {Jarzynski relation (JR)}

The Jarzynski relation \cite{jarz97} 
\beq\langle \exp[-w/T]\rangle =\langle \exp[-\Delta \F/T]\rangle 
\label{eq:JR}
\ee
expresses the free energy difference $\Delta F\equiv \F(\lt)-\F(\lo)$
between two equilibrium states as
a non-linear average over the work required to drive the system from one
equilibrium state to another. In the present formalism it follows 
for motion in a time-dependent potential $V(x,\la)$  without any
additional  non-conservative
force $f\equiv 0$  by plugging into the general FT (\ref{eq:IFT2}) 
the  equilibrium initial distribution
$p_0(x)=\exp[-(V(x,\lo)-\F(\lo))/T]$  and the function $p_1(x) =
\exp[-(V(x,\lt)-\F(\lt))/T]$. It should
be clear that this  choice for $ p_1(x)$ exploits  a mathematical freedom.
It
does not imply that the system has acquired this new equilibrium distribution
at the end of the process. In fact, the actual distribution at the end
will be $p(x,t)$.

\subsection {Bochkov-Kuzovlev relation (BKR)}

For a system initially in equilibrium in a 
time-independent potential $V_0(x)$,
which is for $0\leq \tau\leq t$ subject to an additional space and
time-dependent force $f(x,\tau)$, one obtains from (\ref{eq:IFT2})
 the relation
\beq
\langle \exp[-\tilde w/T]\rangle = 1
\label{eq:bk}
\ee
with
\beq
\tilde w\equiv \int_{x_0}^{x_t} f(x,\la) dx 
\ee by choosing
$p_1(x)=p_0(x)=\exp[-(V_0(x)-\F_0)/T]$.
Under these conditions, 
the exponent $\tilde w$ is the work performed at the system. Since this
 relation derived  much earlier by Bochkov and Kuzovlev \cite{boch81,boch81a} 
looks
almost like the Jarzynski relation there have been both  claims that the
two are the same and some confusion around the apparent contradiction
that $\exp[-w/T]$ seems to be both  $\exp[-\Delta \F/T]$) or 1.

 The present derivation shows that the two relations are different
since they apply a priori to somewhat different situations. 
The JR as discussed above
applies to processes in a time-dependent potential, whereas the BKR
relation as discussed here applies to a process in a constant potential
with some additional force. If, however,  in the latter case, 
this explicit force
arises from a potential as well, $f(x,\tau) = -  V_1'(x,\tau),$ there 
still seems
to be an ambiguity. It can be resolved by reckognizing that in this 
case
the work entering the BKR (\ref{eq:bk}) 
\beq \tilde w = \int dx f =- \int dx V_1'(x) = -\Delta V_1
+ w
\ee 
 differs by a boundary term
from the definition of work $w$ given in
eq. (\ref{eq:work}) and
used throughout this paper. 
Thus, if the force arises from a time-dependent but
conservative potential both the BKR in the form 
$\langle \exp[-\tilde w/T]
\rangle = 1$ and the JR (\ref{eq:JR}) hold. 
The connection between the two relations has previously
 been discussed within a Hamiltonian dynamics approach \cite{jarz07}.

\subsection{ Hummer and Szabo relation}

As an important application, the Jarzynski relation can been used to
reconstruct the free energy landscape of a biomolecule
 $G(x)$ where $x$ denotes a ``reaction coordinate'' like the end-to-end
distance in forced protein folding as reviewed in \cite{rito06}. 
If one end of such a protein is
attached to an AFM tip or an optical trap via a potential $(k/2)(x-\la)^2$  
where $k$ is the stiffness of the cantilever or trap and $\la$ the
time-dependent center, the total potential reads
\beq
V(x,\la) = G(x) +(k/2) (\la -x)^2  .
\ee
The initial distribution $p_0(x)=\exp[-(V(x,\lo)-\F(\lo))/T]$ and  the
choice
\beq
p_1(x)=\delta (x-z)\exp-[V(x,\lt)/T]/\exp-[V(z,\lt)/T]
\ee 
plugged into (\ref{eq:IFT}) leads to the
potential reconstruction formula 
\beq
e^{-G(z)} = \langle \delta[z-x(t)]e^{-w(t)}\rangle ~~e^{(k/2)(z-\l(t))^2}
e^{\F(\lo)}
\ee
first derived by Hummer and Szabo \cite{humm01} 
using a Feynman-Kac approach.
Thus to get the potential $G(z)$ it is sufficient to select those
trajectories that
have reached $z$ after time $t$ and record the corresponding
 work  $w(t)=k\int_0^t d\tau(\la-x(\tau))$ 
 accumulated up to time $t$. An experiment on unfolding
RNA has been one of the first real-world test of this $z$-resolved
 Jarzynski relation \cite{liph02}.

\subsection {Integral ``end-point'' relations}

A variety of ``end-point'' relations can be generated 
from (\ref{eq:IFT}) as follows.

By choosing $p_1(x)=p(x,t)g(x)/\langle g(x_t) \rangle $, one obtains
\beq
\langle g(x_t)\exp[-\Delta s\t]\rangle = \langle g(x_t) \rangle  
\label{eq:e1}
\ee for any function $g(x)$ \cite{schm06a}.
Likewise, for $f\equiv 0$ and $V(x,\la)$, by choosing
$p_1(x)=g(x)\exp[-V(x,\lt)-\F(\lt)]/\langle g(x)\rangle_{{\rm eq}, \lt}$,
one obtains
\beq
\langle g(x_t)\exp[-(w-\Delta \F)/T]\rangle =  
\langle g(x)\rangle_{{\rm eq}, \lt}
\label{eq:e2}
\ee
which has been first derived by Crooks \cite{croo00}. Here, the average
on the right hand side is the equilibrium average at the final
value of the
control parameter.

In the same fashion, one can derive
\beq
\langle  g(x_t)\exp[-\tilde w /T]\rangle =  
\langle g(x)\rangle_{{\rm eq}, \lo}
\label{eq:e3}
\ee
by choosing $ p_1(x) = g(x)\exp[-V(x,\lo)-\F(\lo)]/\langle g(x)\rangle_{{\rm
    eq}, \lo}
$
for a time-independent potential and arbitrary force $f(x,\tau)$ which
is the end-point relation corresponding  to the BKR.

\section {Non-equilibrium steady states (NESSs)}

\subsection {Characterization} 
Non-equilibrium does not necessarily require that  the system is driven
by
time-dependent potentials or forces as discussed so far. A non-equilibrium steady
state (NESS) is generated if time-independent but
non-conservative forces $f(x)$ act on the system. Such systems are characterized by a 
time-independent or stationary distribution 
\beq
p^s(x)\equiv \exp[-\phi(x)]  .\ee
As a  fundamental difficulty, there seems to be no simple way to
calculate $p^s(x)$ or, equivalently, the
 ``non-equilibrium potential'' $\phi(x)$. In one dimension, 
it follows from quadratures but for more degrees of freedom, setting the
right-hand-side of the Fokker-Plank equation (\ref{eq:fp}) to zero represents a 
formidable
partial differential equation. Physically, the complexity arises from
the presence of a non-zero 
stationary current (in the full configuration space)
\beq
j^s(x)=\mu F(x) p^s(x)-D\partial_x p^s(x) \equiv v^s(x)p^s(x)
\ee
which defines the mean local velocity $v^s(x)$. This current leads to
a mean entropy production rate
\beq
\sigma\equiv \langle\Delta s\t\rangle /t = \int dx j^s D^{-1} j^s/p^s .
\ee
Even though the stationary distribution  and currents can not be
calculated in general, a variety of exact relations concerning entropy
production
and heat dissipation have been derived.

\subsection{Detailed fluctuation theorem (DFT)}

In a NESS, 
the (detailed) fluctuation theorem 
\beq
\pmst/\pst=\exp[-\Delta s\t]
\label{eq:dft}
\ee
expresses a symmetry which the probability
distribution $\pst$ for the total entropy production after time $t$ in the
steady state obeys. This relation has first been found in simulations of
two-dimensional sheared fluids \cite{evan93} 
and then been proven by Gallavotti and
Cohen \cite{gall95} using assumptions about chaotic dynamics. A much simpler proof has
later
been given by Kurchan \cite{kurc98} and Lebowitz and Spohn \cite{lebo99}
using a stochastic dynamics
for diffusive motion. Strictly speaking, in all these works the relation
holds only asymptotically in the long-time limit since entropy production
 had been associated with what is here called entropy production in the
medium. 
If one includes the entropy change of the system (\ref{eq:s}), 
the
DFT holds even for finite times in the steady state \cite{seif05a}. 
This fact shows again
the advantage of the entropy definition introduced above.

While the DFT for (medium) entropy production has been tested experimentally 
for quite
a number of systems,  a first test including the system entropy 
has recently
been achieved for a colloidal particle driven by a constant force
along a periodic potential
\cite {spec07}. This experimental set-up constitutes the simplest realization
of genuine NESS. The same set-up has been used  to test other recent aspects
of 
stochastic
thermodynamics like the possibility to infer the potential $V(x)$ from
the measured stationary distribution and current  \cite{blic07a}  
or a generalization of the
Einstein relation beyond the linear response regime \cite{spec06,blic07}.

The DFT for total entropy production holds even
under the more general situation of periodic driving 
$F(x,\tau)=F(x,\tau + \tau_p)$, where $\tau_p$ is the period, if (i)
the system has settled into a periodic distribution
$p(x,\tau)=p(x,\tau + \tau_p)$, and (ii) the trajectory length $t$ is an
integer multiple
of $\tau_p$.

For the distribution of work $p(W)$, a similar DFT can be proven provided
the protocol is symmetric $\la=\l(t-\tau)$, the non-conservative force zero,
and the systems starts in equilibrium initially. For such conditions,
the DFT for work was recently tested experimentally using a
a colloidal particle
 pushed periodically by a laser trap
against a repulsive substrate \cite{blic06}.

\subsection{Transitions between different NESSs}

Just as a system can be driven from one equilibrium state to another by
a time dependent potential $V(x,\la)$, one can generate a transition from
one NESS characterized by $f_1(x)$ to another one at $f_2(x)$ by a 
time-dependent
force $f(\tau)$. For the  thermodynamic analysis of such a 
transition, it has
become convenient to split the dissipated heat quite generally into two
contributions,
\beq 
\Delta q \equiv \Delta q\hk + \Delta q\ex
\ee
where the house keeping heat 
\beq
q\hk\equiv \int_0^td\tau \dot x (\tau) \mu^{-1}v^s(x(\tau))
\label{eq:hk}
\ee
is the heat inevitably dissipated in maintaining a NESS \cite{oono98,hata01}. 
For a transition
from one NESS to another, one has both the IFT for the house
keeping heat \cite{spec05a} 
\beq
\langle \exp[-q\hk/T]\rangle=1
\label{eq:ifthk}
\ee and the Hatano-Sasa relation \cite{hata01}
\beq
\langle \exp{-[q\ex/T +\Delta \phi(x,\l)]}\rangle = 1  .
\label{eq:hs}
\ee  
The latter relation allows one to obtain Clausius-type inequalities for the
minimal entropy change related to transitions between different NESSs.

Experimentally, the Hatano-Sasa relation has been verified for a colloidal particle
pulled through a viscous liquid at different velocities which
corresponds to different steady states \cite{trep04}.

\section {General stochastic dynamics}

\subsection {Entropy production}

For the mechanically driven systems described so far the identification of
a first law is simple since both internal energy and applied work are
rather clear concepts. On the other hand the proof of both the
IFT and the DFT shows that the first law does not crucially enter. In
fact, the proof of these theorems exploits only the fact that under 
time-reversal entropy production changes sign. 
If the changes in the medium entropy are written in the form (\ref{eq:heat},
\ref{eq:dsm}) only 
transition rates enter. Likewise, the identificiation of system entropy
requires only a solution of the Fokker-Plank equation. 
Therefore, similar relations should hold for a much larger class of 
stochastic dynamic models without reference to a first law.

Indeed, one can derive these relations for  stochastic dynamics on an
arbitrary set of states $\{n\}$ where transitions
between states $m$ and $n$ occur with a rate $w_{mn}(\l)$, which may
depend on an externally controlled time-dependent parameter $\l(\tau)$. The
master equation for the time-dependent probability $p_n(\tau)$ then reads
\begin{equation}
  \partial_\tau p_n(\tau) = \sum_{m \not = n}
  [w_{mn}(\l) p_m(\tau) - w_{nm}(\l) p_n(\tau)].
\label{eq:me}
\end{equation}

The analogue of the fluctuating trajectory $x(\tau)$ in the mechanical case
becomes a stochastic trajectory $n(\tau)$ that starts at $n_0$
and jumps at times $\tau_j$ from $n_j^-$ to
$n_j^+$ ending up at $n_t$.
As entropy along this trajectory, one defines \cite{seif05a}
\beq
s(\tau)\equiv - \ln \pnt 
\ee
where $\pnt$ is the solution  $p_n(\tau)$ of the master equation (\ref{eq:me})
for a given
initial distribution $p_n(0)$ taken along the specific trajectory $n(\tau)$.
As above, this entropy will depend on the chosen
initial distribution.

The entropy $s(\tau)$ becomes time-dependent due to two sources.
First, even if the system does not jump, $\pnt$ can be time-dependent
 either for time-independent rates due to possible
relaxation from a non-stationary
initial state or, for time-dependent rates, due to the explicit
time-dependence of $p_n(\tau)$. Including the jumps, the change of system
entropy reads
\begin{eqnarray}
\dot s(\tau)&=& -{\partial_\tau \pnt\over
\pnt} -\sum_j\delta(\tau-\tau_j)\ln{\pjp\over \pjm}\\
&\equiv& \dst-\dsm .
\end{eqnarray}
where we define the change in medium entropy to be
\beq
\dsm\equiv
 \sum_j\delta(\tau-\tau_j)
\ln{\wmp\over\wpm}  
\label{eq:sm}  .
\ee

For a general system, associating the
logarithm of the ratio between forward jump rate and backward jump rate
with an entropy change of the medium seems to be an aritrary definition. 
Two facts
motivate this choice. First, it corresponds precisely to what has been
identified as exchanged heat in the mechanically driven case
(\ref{eq:heat},\ref{eq:dsm}). Second, 
one can easily show  \cite{seif05a} that with this choice 
the
total entropy production $\Delta s\t$ fulfills both the IFT 
(\ref{eq:IFT}) for
arbitrary initial condition, arbitrary driving and any length of
trajectory and the DFT (\ref{eq:dft}) in the steady state, i.e. for constant
rates. Of course, in a general system, there is no justification
to identify the change in medium entropy with an exchanged heat.

\subsection{Two classes of systems}
 
For a discussion, it is appropriate 
to distinguish two classes of systems based on whether or not their
stationary distribution $p^s_n$ for fixed 
 $\l$ obeys the detailed balance condition
\beq
p_n^s(\l)w_{nm}(\l)=p_m^s(\l)w_{mn}(\l) .
\label{eq:db}
\ee

\subsubsection{ Detailed balanced systems}
System which obey  detailed balance formally resemble  mechanically
driven systems without non-conservative force  since for the
latter, at fixed potential,   detailed balance holds  as well. Exploiting
this
analogy,  one can assign a (dimensionless) internal ``energy'' 
\beq
\epsilon_n(\l)\equiv -  \ln p^s_n(\l)
\ee 
to each state.
The ratio of the rates then obeys
\beq
{\wnm(\l)\over\wmn(\l)} = \exp[\epsilon_n(\l)-\epsilon_m(\l)]
\ee
which looks like  the familiar detailed balance condition in equilibrium.
For time-dependent rates $\wnm(\la)$, one can now  
formally associate a first law
along the trajectory as follows.
The analogue of the work is written in the form
\begin{eqnarray}
w&\equiv& \int_0^t \partial_\l \epsilon_n(\l(t))= \sum_j\ln{\wmp\over \wpm}
 +\epsilon_{n(t)}(\lt)-
\epsilon_{n(0)}(\lo)\nonumber\\
&\equiv& \Delta q + \Delta V
\end{eqnarray}
where the second lines identifies the ``heat'' with the medium entropy change and
the last term represents a change in ``internal energy''.
Even though one should not put physical meaning into these definitions
for an abstract stochastic dynamics, the analogy helps to see immediately
that the fluctuation relations
quoted above for zero non-conservative force hold for these\
more general systems as well.
Specifically, one has the ``generalized'' JR (\ref{eq:JR}) and 
the corresponding end-point relation (\ref{eq:e2})
  with $\Delta\F = 0$ and $T = 1$. Similarly, 
with the identification
\beq
\tilde w\equiv
\sum_j[\eps_{\njp}(\tau_j)-\eps_{\njm}(\tau_j)]-[\eps_{\njp}(0)-\eps_{\njm}(0)]
\ee 
both  (\ref{eq:bk}) and (\ref{eq:e3}) with $T= 1$ hold for such a master
equation dynamics. The initial state in all cases is the steady state 
corresponding to $\lo$.

The simplest realization of such a detailed-balanced 
system is any two-level system
with time-dependent rates. In recent experiments  \cite{schu05,tiet06}, 
an optically driven
defect center in diamond has been used to test the IFT for total
entropy production and the analogy of the Jarzynski relation for such a
general stochastic dynamics \cite{seif04}.

\subsubsection {Unbalanced systems}

The more interesting class of networks, however, are those for
which at constant $\l$ the detailed balance condition (\ref{eq:db})
is not fulfilled. Within the mechanical analogy,
 they correspond to 
systems with non-vanishing non-conservative force $f\not = 0$. 
For a mapping to this
case, one would have  to find the analogy of the conservative potential and the
force $f$ which is responsible for breaking detailed balance. The apparent
arbitrariness of such a division is resolved by forming the analogy of the
house-keeping heat (\ref{eq:hk}) as
\beq
q\hk[n(\tau)] \equiv \sum_j\ln{\pjm^s\wmp\over\pjp^s\wpm} .
\ee 
For detailed balanced systems, this expression vanishes identically. With
this identification, it is clear that the IFT for the
housekeeping heat (\ref{eq:ifthk})
 and the Hatano-Sasa relation (\ref{eq:hs}) hold as well using
these definitions (and $T=1$). For further fluctuation relations in such systems, see the
comprehensive review \cite{harr07}.

\section{(Bio)chemically driven systems}

Biochemical reactions constitute another 
 important class of systems for which an embedding heat bath provides the
source of  stochastic dynamics. Therefore,
the concepts reviewed so far are ideally suited to develop a \sd of such 
systems. Conceptually, two situations should be distinguished described here
only briefly.

\subsection {A single enzym or motor}
An enzym or molecular motor can be considered as a system which stochastically
undergoes  
transitions from one state $m$ to another state $n$. In such a 
transition, a chemical reaction may be involved like hydrolysis which
transforms
one molecule ATP to ADP and a phosphate. These three molecular
species are externally
maintained
at non-equilibrium conditions thereby providing a source of chemical energy,
i.e., chemical work, to the system. In each transition, this work will  
be transformed into mechanical work, dissipated heat, or changes in the 
internal energy (with any combination of positive and negative 
contributions).  The formalism of \sd allows to identify work, heat, and
internal
energy for each single transition \cite{schm06a,seif05}
 in close analogy to the mechanically driven  case,
thus illustrating the concepts described above for a more general stochastic 
dynamics. 

\subsection {Chemical reaction networks}

Chemical reaction networks consist of a number of possible reactions taking place
between different types of reactants. The rates for each reaction are proportional
to the number of possible reaction partners thus involving the stochiometric
coefficients. In order to describe such a network
under non-equilibrium conditions, one should
separate the reactants into two classes. The first class comprise species
whose concentration, i.e.,  chemical potentials, are fixed externally at
constant values (like the ATP concentration in the cell). The second species
comprises molecules whose number is traced along a trajectory. Under such
conditions,  chemical work, dissipated heat and changes in iternal
energy can be identified on the level of a single reaction trajectory
for which the general formalism yields IFTs, 
generalization of the Jarzynski relation and other relations 
\cite{schm06}. So far, no experiments illustrating these 
concepts
seem to have been reported.

\section{Stochastic dynamics of fields}

\def\ph{\phi(r,\tau)}
\def\phso{\phi(r',\tau)}
\def\phs{\phi(r',\tau')}
\def\ze{\zeta(r,\tau)}
\def\zes{\zeta(r',\tau')}
\def\ps{p^s(\phi(r,\tau))}
\def\pso{p(\phi(r,\tau),\tau)}
\def\h{h(r,\tau)}

The Langevin dynamics discussed for the mechanically driven case
can easily be generalized to $N$ coupled degrees of freedom
\cite{schm06a,cher06}. A further generalization to the dynamics of stochastic
fields, which as an exciting perspective looks obvious, seems
not yet to   been explored 
systematically.
 
We assume a dynamical equation for a field $\ph$ with $r\in \mathbb{R}^d$ of the type
\beq
\partial_\tau  \ph = F[\ph] + \ze
\ee
where $F[\ph]$ is an arbitrary possibly non-linear functional
and $\ze$  Gaussian white noise of strength
\beq
\langle \ze \zes\rangle = 2 D(r-r') \delta (\tau-\tau')
\ee
with a spatial correlation $D(r-r')$. Motivated by the mechanically driven 
case, one  defines
a change in medium entropy 
\beq 
\Delta s_{\rm m} \equiv \int_0^t d\tau \int dr\int dr' ~\partial_\tau \ph
D^{-1}(r-r')   F[\phso]   .
\ee
and a change in entropy of the field
\beq
\Delta s \equiv -\ln\pso|^t_0 
\ee
where $\pso$ is the evolving probability distribution for the field.

The validity of the IFT for total entropy production
 (\ref{eq:IFT}) for arbitrary
initial state and arbitrary length $t$ of the trajectory is obvious.
If the dynamics settles into a steady state $\ps$, 
the DFT (\ref{eq:dft}) also holds.

As an illustrative example consider the KPZ equation 
\beq
\partial_\tau h=\l (\nabla h)^2 + \nu \nabla^2 h + \zeta
\ee
for the height field $\h$ of a growing interface where $\l$ measures
the strength of the non-linearity and $\nu$ corresponds to a 
surface tension \cite{kardar}.  The white noise is uncorrated in space and time
$\langle \ze\zes\rangle = 2 \delta (r-r')\delta(\tau-\tau')$
(setting $D$ to 1 for notational simplicity).

In the steady state, the total entropy production becomes
\begin{eqnarray}
\Delta s\t =&~& \l \int_0^t d\tau\int dr  (\partial_\tau h)(\nabla h)^2
\nonumber \\
&-& \left[(\nu/2) \int dr (\nabla h)^2 + \ln p^s(h)\right]^t_0 .
\label{eq:fdtkpz}
\end{eqnarray}
In one dimension, the known stationary distribution
$
p^s(h) \sim \exp[-(\nu/2)\int dr ~  (\nabla h)^2]
$ cancels
the boundary term of the medium entropy so that the total entropy
production over a time $t$ along a trajectory in the steady state
is given exactly by
\beq
\Delta s\t =  \l \int_0^td\tau \int dr  (\partial_\tau  h)(\nabla h)^2 .
\ee

The DFT (\ref{eq:dft}) now makes a statement about the probability 
distribution of the time integral of this particular product involving
three powers of the field. By expanding the exponent, 
one can generate relations between particular three point, six point, 
nine point functions
and so on. While the immediate benefit of such  relations is not obvious,
it looks worthwile to explore the consequences of this symmetry
for correlations functions.

In higher dimensions, the stationary distribution is unknown and the
exact statement holds only for the sum of the three terms in (\ref{eq:fdtkpz}). 
Clearly, only the term in the first line is extensive in time so one might 
expect that
focussing on this term in the long time limit may be admissible
but one should be aware of corrections arising from the two boundary terms.
In any case,  one could speculate whether
the DFT restricts exponents or scaling functions if a suitable scaling
ansatz for correlation functions is put into it. 
 
If useful information can be extracted from such an approach for the KPZ
equation, one might afterwards be tempted to look at other stochastic field
equations. A particularly intriguing case are the
Navier Stokes equations
with  noise terms  used to stochastically drive the system on large
scales to model turbulence. In fact, for two-dimensional turbulence such an 
approach has been applied to the enstrophy cascade
\cite{baie05}.

\section {Concluding remark}

The notion ``stochastic thermodynamics'' had been introduced two decades ago
for an interpretation of chemical reaction networks in terms of thermodynamic
notions on the ensemble level \cite{mou86}. From the 
present perspective, it
seems even more appropriate to use this term for the refined description
along the fluctuating trajectory. Both for mechanically and chemically driven
systems in a surrounding heat bath, the thermodynamic concepts can literally
and consistently be applied on this level. As a generalization to arbitrary
stochastic dynamics, analogues of work, heat and internal energy obey similar
exact relations which ultimately all arise from the behaviour of the dynamics
under time-reversal. How much closer such an approach can lead us towards a
systematic understanding of non-equilibrium phenomena in general is a question
too early to be answered yet. 

\section {Acknowledgment}

I thank  T. Schmiedl and T. Speck for stimulating interactions and the latter
for a critical reading of this manuscript; C. Bechinger,
V. Blickle, C. Tietz and J. Wrachtrup for fruitful experimental
collaborations;
and C. van den Broeck, P. H\"anggi, C. Jarzynski, F. Ritort, G. Sch\"utz,
H. Spohn, J. Vollmer, H. Wagner and R. Zia for illuminating discussions.

%
%
%
%

\end{document}